\documentclass[11pt,twoside]{article}
\usepackage{./asp2014}

\aspSuppressVolSlug
\resetcounters

\begin{document}

\title{An internally consistent distance framework in the Local Group}
\author{Richard de Grijs$^{1.2}$ and Giuseppe Bono$^{3,4}$}
\affil{$^1$Kavli Institute for Astronomy \& Astrophysics and
  Department of Astronomy, Peking University, Yi He Yuan Lu 5, Hai
  Dian District, Beijing 100871, China; \email{grijs@pku.edu.cn}}
\affil{$^2$International Space Science Institute--Beijing, 1
  Nanertiao, Zhongguancun, Hai Dian District, Beijing 100190, China}
\affil{$^3$Dipartimento di Fisica, Universit\`a di Roma Tor Vergata,
  via Della Ricerca Scientifica 1, 00133, Roma, Italy;
  \email{Giuseppe.Bono@roma2.infn.it}}
\affil{$^4$INAF, Rome Astronomical Observatory, via Frascati
  33, 00040, Monte Porzio Catone, Italy}

% This section is for ADS Processing.  There must be one line per author.
\paperauthor{Richard de Grijs}{grijs@pku.edu.cn}{0000-0002-7203-5996}{Peking University}{Kavli Institute for Astronomy and Astrophysics}{Beijing}{}{100871}{China}
\paperauthor{Giuseppe Bono}{Giuseppe.Bono@roma2.infn.it}{0000-0002-4896-8841}{Universit\`a di Roma Tor Vergata}{Dipartimento di Fisica}{Rome}{}{00133}{Italy}

\begin{abstract}
Accurate and precise astronomical distance determinations are crucial
for derivations of, among others, the masses and luminosities of a
large variety of distant objects. Astronomical distance determination
has traditionally relied on the concept of a `distance ladder.' Here
we review our recent attempts to establish a highly robust set of
internally consistent distance determinations to Local Group galaxies,
which we recommend as the statistical basis of an improved
extragalactic distance ladder.
\end{abstract}

\section{Establishing the distance ladder}

Distance determination is among the most fundamental tasks of the
astronomer. Accurate and precise distances allow us to derive a
multitude of basic physical parameters of objects as diverse as stars,
galaxies, active galactic nuclei, and even of cosmological large-scale
structures. Yet, to determine the distances to the nearest galaxy
clusters, for instance, we must understand the physics of the galaxies
in the nearby Universe. Distances to galaxies in the Local Group, in
turn, often rely on our detailed understanding of the physics of a
range of stellar tracers. This goes to show that the process of
distance determination to increasingly distant target objects relies
on accurate and precise knowledge of the distances to their more local
counterparts, with each rung of the so-called `distance ladder'
depending for its calibration on equally good or better calibration of
the previous, closer rung (e.g., de Grijs 2011).

Whereas geometric distances can be obtained to a range of objects in
our own Milky Way galaxy and even to the nearest galaxies in the Local
Group, provided that one chooses one's tracers carefully (e.g.,
Pietrzy\'nski et al. 2013), individual tracers do not always provide
robust measurements for the purposes of distance calibration. Yet, the
Local Volume, and the Local Group in particular, represents the single
most important environment to calibrate our distance zero point. 

For at least a century, numerous authors have derived distances to one
or more galaxies in the Local Group (see, e.g., de Grijs et al. 2014;
de Grijs \& Bono 2014, 2015, 2016; and references therein),
characterized by a range in accuracy and precision. Over the course of
the past four years, we have attempted to establish a highly robust
set of internally consistent distance determinations to Local Group
members (de Grijs et al. 2014; de Grijs \& Bono 2014, 2015), including
to the centre of the Milky Way (de Grijs \& Bono 2016), which we hope
can serve as the long-targeted distance zero point and form the basis
of an improved extragalactic distance ladder. As we will see below, we
have opted to tie our distance scale to the distance to the Large
Magellanic Cloud (LMC), for which we have adopted a distance modulus
of $(m-M)_0 = 18.49$ mag (Pietrzy\'nski et al. 2013; de Grijs et
al. 2014; note that since this value is our adopted distance modulus,
we do not include an uncertainty) throughout.

\subsection{Distance to the Galactic Centre}

As our starting point in the Local Group, we take the Milky Way's
centre as our initial benchmark. The distance to the Galactic Centre
is indeed used as a reference value for a variety of methods of
distance determination, both inside and outside the Milky Way. Its
actual value has a direct impact on the distances, masses, and
luminosities one can derive for many Galactic objects, as well as on
the mass, size, and rotation characteristics of the Milky Way as a
whole. Most luminosities and masses scale as (distance$^2$), while
masses based on total densities or orbital modelling scale as
(distance$^3$). Higher-accuracy distance determinations allow for
improved calibration of the zero points of many secondary distance
tracers, including of the often-used Cepheid and RR Lyrae
period--luminosity relations. In turn, this leads to improved
estimates of globular cluster ages, the Hubble constant, and the age
of the Universe. In addition, improved calibration of the local
distance scale carries the potential to yield better constraints on a
range of cosmological scenarios.

We perused the extensive literature on the Galactic Centre and
embarked on a careful analysis of a large number of tracer populations
(de Grijs \& Bono 2016). We included both those populations defining a
`centroid' with the Galactic Centre at its centre (e.g., globular
clusters, red clump stars, as well as Cepheid, RR Lyrae and Mira
variables) and kinematic tracers (including variable stars, open
clusters, maser sources, and objects tracing the Galactic disk). We
thus arrived at a recommended Galactic Centre distance of $R_0 = 8.3
\pm 0.2 \mbox{ (statistical)} \pm 0.4$ (systematic) kpc, corresponding
to $(m-M)_0 = 14.60 \pm 0.05 \mbox{ (statistical)} \pm 0.10$
(systematic) mag.

Compare and contrast this recommendation with the current-best
determinations based on geometric considerations, including the orbits
of the Galactic Centre's `S stars' and statistical parallax
determinations of the central population of stars in the Milky
Way. The S stars have yielded values for the Galactic Centre distance
ranging from $R_0 = 8.0 \pm 0.3$ kpc (Ghez et al. 2008a,b) to $R_0 =
8.2 \pm 0.34$ kpc (Gillessen et al. 2013). Similarly, statistical
parallax measurements resulted in $R_0 = 8.24 \pm 0.12$ kpc
(Rastorguev et al. 2016). By contrast, Matsunaga et al.'s (2011)
determination of $R_0 = 7.9 \pm 0.2$ kpc (based on application of the
period--luminosity relation to three Cepheid variables in the Galactic
nucleus) is systematically smaller.

Separately, we arrived at the conclusion that $R_0 = 8.3$ kpc implies
that the circular rotation speed at the solar circle, $\Theta_0$, as
traced by the Galaxy's mass-bearing components is $\Theta_0 = 225
\mbox{ (statistical)} \pm 3 \pm 10$ (systematic) km s$^{-1}$, so that
$\Theta_0/R_0 = 27.12 \pm 0.39 \mbox{ (statistical)} \pm 1.78$
(systematic) km s$^{-1}$ kpc$^{-1}$ (de Grijs \& Bono 2017).

\subsection{The Large Magellanic Cloud}

The distance to the LMC is truly the first rung on the extragalactic
distance ladder. The galaxy hosts statistically large samples of
so-called `standard candles,' so that it represents a great benchmark
object for distance calibration. Because of its disk-like geometry,
all standard candles are located at roughly the same distance (but
note that depth effects are not completely negligible). In addition,
they are affected by only little extinction. The LMC offers us a great
opportunity to cross calibrate many tracers simultaneously, and
possibly even link them to their Galactic counterparts.

It also hosts a number of geometric distance tracers, including the
supernova (SN) 1987A and a good number of eclipsing binary
systems. The systematic uncertainty in the LMC distance was singled
out by the {\sl Hubble Space Telescope} ({\sl HST}) Key Project
(HSTKP) on the Extragalactic Distance Scale (Freedman et al. 2001) as
being of key importance for determining the precision of the Hubble
constant, $H_0$, and until recently contributed most of the remaining
systematic uncertainty. The HSTKP team determined $H_0 = 72 \pm 3
\mbox{ (statistical)} \pm 7$ (systematic) km s$^{-1}$ Mpc$^{-1}$.

En passant, the HSTKP team determined a distance modulus for the LMC,
$(m-M)_0 = 18.50 \pm 0.10$ mag, corresponding to $D_{\rm LMC} =
50.1^{+1.4}_{-1.2}$ kpc. Freedman et al. (2001) based these
determinations on a revised calibration of the Cepheid
period--luminosity relation (adopting the maser-based distance to NGC
4258 as their distance calibration) and numerous secondary
techniques. At the same time, this result finally resolved the
long-standing `short' versus `long' LMC distance debate, which had
been raging in the community for a number of decades.

However, trends in subsequent LMC distance determinations were
questioned by Schaefer (2008), who claimed that ``all 31 measurements
since 2001 seem to cluster too tightly around the {\sl HST} Key
Project's value.'' Schaefer (2008) attributed his findings to the
presence of `publication bias' or a potential bandwagon effect. In de
Grijs et al. (2014) we set out to redo this latter analysis based on
the most comprehensive data mining approach of the literature carried
out until that time. 

We concluded that we could not support Schaefer's (2008) claims. For
instance, his suggestion that the uncertainties in the post-2002
measures were not distributed like a Gaussian assumes that the
pre-2002 uncertainties were Gaussian-like. They are not. This also
assumes that conditions have remained comparable, which again is not
justified. Schaefer (2008) did not undertake a detailed assessment of
the systematic uncertainties, while his statistical tests were in
essence based on application of the Kolmogorov--Smirnov test. This
latter test assumes a Gaussian distribution of LMC distance
measurements, as well as a sample of independent and identically
distributed values. Neither assumption is applicable to the body of
LMC distance measurements. By virtue of our significantly enlarged
database of LMC distance measurements, which avoided the pitfalls
associated with the earlier incomplete database where the presence of
gaps was found to hide correlations, we found that we could not
conclude that publication bias significantly affected the published
LMC distance determinations.

Having perused more than 16,000 articles, we eventually collected 233
post-1990 derivations of the LMC distance modulus. In our subsequent
analysis, we had to make a number of choices and assumptions. For
multiple measurements published in the same paper, we included them
separately if they comprised `final' results (after all, any
differences show the effects of systematic uncertainties). We
therefore did not only consider weighted means. If correlated results
were provided in a given paper, we considered these representative of
the systematic uncertainties. Such effects included variations in
extinction corrections, metallicities or $\alpha$ abundances, and $p$
(`projection') factors. In some papers, the same tracer populations
were used but applied at different wavelengths or they adopted
different calibration methods. Again, this provided us with a handle
on the systematic uncertainties; only 47 articles included their own
assessments of these systematic effects.

Systematic uncertainties encountered in our analysis included
differences in zero-point calibrations (e.g., {\sl Hipparcos}, {\sl
  HST}/Fine Guidance Sensor, or ground-based, interferometric
parallaxes), as well as differences in Baade--Wesselink analyses; the
functional form of the calibration relations adopted;
Lutz--Kelker-type biases in parallax measurements; differences in the
metallicity scale and extinction corrections; transformations between
filter systems (e.g., {\sl HST} versus ground-based $UBVRI$); and the
location of the emission from SN 1987A as a function of wavelength.

In terms of parallax measurements, observational uncertainties cause
objects which are in reality located outside the adopted lower
parallax limit to scatter into the sample's selection volume and vice
versa. Since there are more objects just outside than just inside the
selection boundary (at least for predominantly uniformly distributed
objects), more objects will be scattered into than out of the sample,
so that a systematic bias is introduced. This so-called `Lutz--Kelker
bias' applies to parallax measurements of any {\it sample} of objects;
it does not strictly apply to {\it individual} parallaxes.

As regards the appearance of SN 1987A, the key questions requiring
exploration include, Does the emission used to measure delay times at
a variety of wavelengths actually originate from the same region(s) in
the ring? Does emission start immediately when photons hit the gas?
Is there a delay? It has been suggested that ultraviolet lines, such
as N{\sc iii}]/N{\sc iv}] may originate from the ring's inner edge,
    while the optical [O{\sc iii}] lines may have come from the main body.
Using the proper geometry, including a finite ring thickness, the
ultraviolet light curve could then result in an underestimate of the
light travel time across the optical ring diameter (and thus in the
distance) of up to 7\%. This scenario is unlikely, however, given the
very similar ionisation potentials of [O{\sc iii}] and N{\sc
iii}]/N{\sc iv}], as well as their likely spatial distributions
(e.g., Gould \& Uza 1998; and references therein).

Following a careful assessment of the statistical and systematic
uncertainties associated with all reported post-1990 LMC distance
measurements, our final recommendation for the LMC's distance modulus
is $(m-M)_0 = 18.49 \pm 0.09$ mag. This has since been confirmed on
the basis of independent statistical analysis (Crandall \& Ratra
2015). It is also fully consistent with the large body of
Cepheid-based distance measurements to the LMC, which resulted in
$(m-M)_0 = 18.48 \pm 0.08$ mag (de Grijs et al. 2014). We note that
the independent eclipsing binary distance modulus of Pietrzy\'nski et
al. (2013) provides excellent confirmation: $(m-M)_0 = 18.493 \pm
0.008 \mbox{ (statistical)} \pm 0.047$ (systematic) mag, corresponding
to a distance $D_{\rm LMC} = 49.97 \pm 0.19 \mbox{ (statistical)} \pm
1.11$ (systematic) kpc.

\section{Beyond the Large Magellanic Cloud}

We took a similar approach to derive an independent statistically
justified distance modulus to the Small Magellanic Cloud. Although we
noticed some tension at the $2\sigma$ level among the various tracers
used in the literature, our final recommendation, $(m-M)_0^{\rm SMC} =
18.96 \pm 0.02$ mag (statistical uncertainty only) is adequately
supported (within the uncertainties) by the majority of distance
tracers.

For good measure, we proceeded to obtain statistical distances using
the same data mining approach for M31, M33, and a number of smaller
galaxies in their immediate vicinity. Where necessary for relative
distance measurements, we used $(m-M)_0^{\rm LMC} = 18.49$ mag as our
local benchmark for calibration purposes.

Our full set of Local Group distance measurements is included Table 1,
which we offer as our final recommendations. It represents a summary
of having perused tens of thousands of articles in the literature,
eventually leading to a collection of statistically carefully
justified local benchmark distances. In turn, these are meant to
provide a useful cross calibration for {\sl Gaia} and Large Synoptic
Survey Telescope benchmarking.

\begin{table}[!ht]
\caption{Recommended benchmark distance moduli in the Local Group}
\smallskip
\begin{center}
{\small
\begin{tabular}{lc}  % l = left, c = centered
\tableline
\noalign{\smallskip}
\multicolumn{1}{c}{Benchmark} & $(m-M)_0^{\rm rec.}$ (mag) \\
\noalign{\smallskip}
\tableline
\noalign{\smallskip}
Galactic Centre & $14.60 \pm 0.11$ \\
Large Magellanic Cloud & $18.49 \pm 0.09$ \\
Small Magellanic Cloud & $18.96 \pm 0.02$ \\
NGC 185  & $24.00 \pm 0.12$ \\
NGC 147  & $24.11 \pm 0.11$ \\
IC 1613  & $24.34 \pm 0.05$ \\
IC 10    & $24.36 \pm 0.45$ \\
M32      & $24.43 \pm 0.07$ \\
M31      & $24.45 \pm 0.10$ \\
NGC 205  & $24.56 \pm 0.15$ \\
M33      & $24.67 \pm 0.07$ \\
\noalign{\smallskip}
\tableline\
\end{tabular}
}
\end{center}
\end{table}


\begin{thebibliography}{}

\bibitem[Crandall \& Ratra(2015)]{2015ApJ...815...87C} Crandall, S.,
  \& Ratra, B.\ 2015, \apj, 815, 87

\bibitem[de Grijs(2011)]{RdG2011} de Grijs, R.\ 2011, An Introduction
  to Distance Measurement in Astronomy, Wiley Acad. Publ.

\bibitem[de Grijs \& Bono(2014)]{2014AJ....148...17D} de Grijs, R., \&
  Bono, G.\ 2014, \aj, 148, 17

\bibitem[de Grijs \& Bono(2015)]{2015AJ....149..179D} de Grijs, R., \&
  Bono, G.\ 2015, \aj, 149, 179

\bibitem[de Grijs \& Bono(2016)]{2016ApJS..227....5D} de Grijs, R., \&
  Bono, G.\ 2016, \apjs, 227, 5

\bibitem[de Grijs \& Bono(2017)]{2017ApJS..232...22D} de Grijs, R., \&
  Bono, G.\ 2017, \apjs, 232, 22

\bibitem[de Grijs et al.(2014)]{2014AJ....147..122D} de Grijs, R.,
  Wicker, J.~E., \& Bono, G.\ 2014, \aj, 147, 122

\bibitem[Freedman et al.(2001)]{2001ApJ...553...47F} Freedman, W.~L.,
  Madore, B.~F., Gibson, B.~K., et al.\ 2001, \apj, 553, 47

\bibitem[Ghez et al.(2008a)]{2008IAUS..248...52G} Ghez, A.~M., Salim,
  S., Weinberg, N., et al.\ 2008a, IAU Symp., 248, 52

\bibitem[Ghez et al.(2008b)]{2008ApJ...689.1044G} Ghez, A.~M., Salim,
  S., Weinberg, N., et al.\ 2008b, \apj, 689, 1044

\bibitem[Gillessen et al.(2013)]{2013IAUS..289...29G} Gillessen, S.,
  Eisenhauer, F., Fritz, T.~K., et al.\ 2013, IAU Symp., 289, 29

\bibitem[Gould \& Uza(1998)]{1998ApJ...494..118G} Gould, A., \& Uza,
  O.\ 1998, \apj, 494, 118

\bibitem[Matsunaga et al.(2011)]{2011Natur.477..188M} Matsunaga, N.,
  Kawadu, T., Nishiyama, S., et al.\ 2011, \nat, 477, 188

\bibitem[Pietrzy{\'n}ski et al.(2013)]{2013Natur.495...76P}
  Pietrzy{\'n}ski, G., Graczyk, D., Gieren, W., et al.\ 2013, \nat,
  495, 76

\bibitem[Rastorguev et al.(2016)]{Rastorguev16} Rastorguev, A.~S.,
  Utkin, N.~D., Zabolotskikh, M.~V., et al.\ 2016, arXiv:1603.09124

\bibitem[Schaefer(2008)]{2008AJ....135..112S} Schaefer, B.~E.\ 2008,
  \aj, 135, 112

\end{thebibliography}
\end{document}